\documentclass[a4paper]{jpconf}
\usepackage{graphicx}
\begin{document}
\title{Reconstructed Jets at RHIC}

\author{Sevil Salur}

\address{Department of Physics,
University of California, 
One Shields Avenue, Davis, CA 95616}

\ead{salur@physics.ucdavis.edu}

\begin{abstract}
To precisely measure jets over a large background such as pile up in high luminosity p+p collisions at LHC, a new generation of jet reconstruction algorithms is developed. These algorithms are also applicable to reconstruct jets in the heavy ion environment where large event multiplicities are produced. Energy loss in the medium created in heavy ion collisions are already observed indirectly via inclusive hadron distributions and di-hadron correlations.  Jets can be used to study this energy loss in detail with reduced biases.  We review the latest results on jet-medium interactions as seen in A+A collisions at RHIC, focusing on the recent progress on jet reconstruction in heavy ion collisions. 
\end{abstract}

\section{Introduction}

Inclusive hadron distributions and di-hadron correlations at high transverse momentum were utilized to measure jet quenching at RHIC.  While these measurements  are used to avoid the complex backgrounds of high multiplicity heavy ion environments they are biased towards the jet fragmentation particles that has the least interaction with the medium.  Full jet reconstruction in A+A collisions can overcome  theses geometric biases as the energy flow is measured independently of the fragmentation details.  New observables such as energy flow, jet substructure and fragmentation functions that can be measured in multiple channels (inclusive, di-jets, h-jets and gamma-jets) and these measurements is expected to  significantly reduce the effect of the geometric biases.  In this article we review the current status of the new techniques developed for full jet reconstruction in heavy ion collisions at RHIC.  Further experimental details of jet reconstruction measurements utilizing STAR and PHENIX experiments can be found in these proceedings \cite{yuiwwnd, brunawwnd, antwwnd} and from the earlier publications \cite{salurwwnd, salurwwnd2, me, jor, bruna, brunawwnd25, helen, kapitan, ploskon, yuiQM, yuidnp,salurqm,salurdnp}.

\section{Jet Measurements} 

Jet reconstruction algorithms are constantly improved during the last 20 years.  Experimentally and theoretically consistent jet definitions are used to search for jets in leptonic and hadronic colliders and to calculate their expected cross-sections.  Cone, sequential recombination and gaussian filtering algorithms are explored at RHIC.  For an overview of jet algorithms in high energy collisions, see  \cite{jetsref,seymor,catchment,jets,kt,ktref,blazey,gauss,gauss1} and the references therein.

\begin{figure}[h]
\centering
\begin{minipage}{17.5pc}
\includegraphics[width=17.5pc]{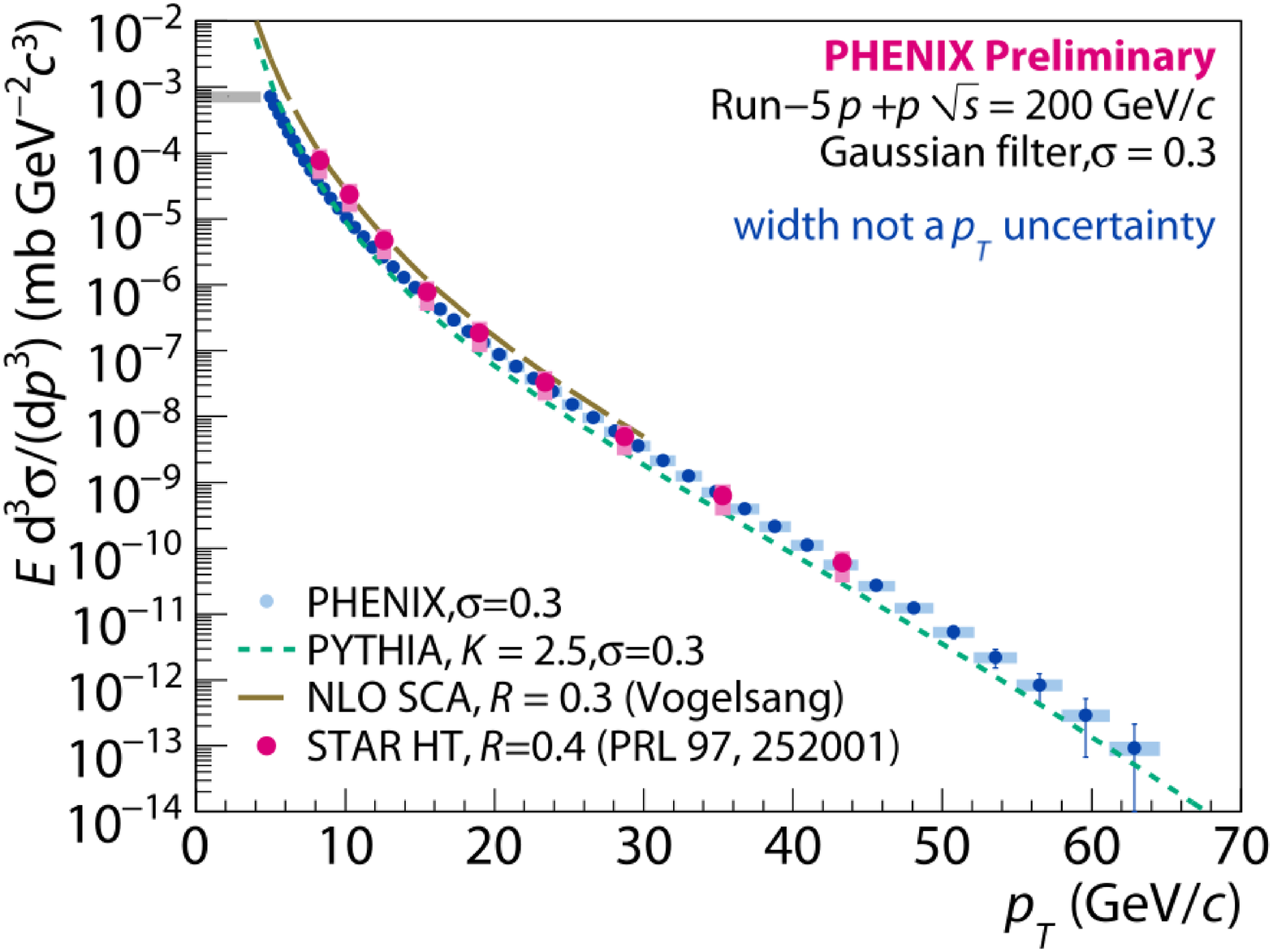}
\caption{\label{fig:ppphenix}Jet cross-section in p+p collisions at  $\sqrt{s}=200$ GeV in the PHENIX detector compared with the jet measurement by the STAR detector \cite{starpp,yuiQM}. Pythia jets and next to the leading order calculations are shown in dashed and solid lines \cite{pythia,phenixnlo}. }
\end{minipage}\hspace{2pc}%
\begin{minipage}{15.5pc}

\includegraphics[width=15.5pc]{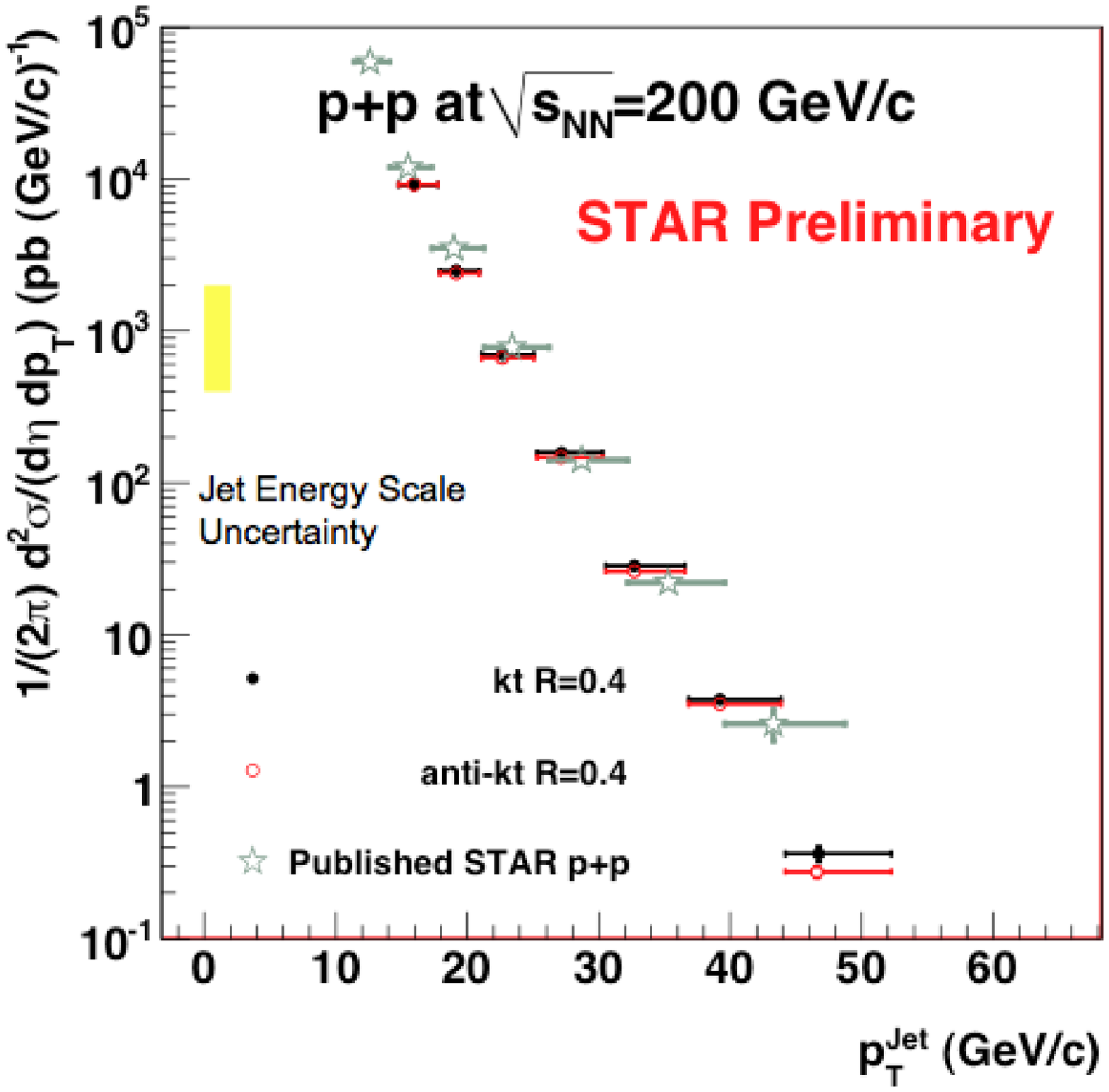}
\caption{\label{fig:pp}Comparison of the jet cross-sections for the p+p collisions obtained by the sequential recombination ($\rm k_{T}$ and anti-$\rm k_{T}$ shown as circles) algorithms and the cone algorithm shown as stars from STAR \cite{ploskon,starpp}.}

\end{minipage} 
\end{figure}

Figures~\ref{fig:ppphenix} and ~\ref{fig:pp} show the inclusive differential cross sections for $p+p\rightarrow jet +X$ at $\sqrt{s}=$200 GeV versus jet $p_{T}$ for both the PHENIX and STAR experiments. Jets that are shown in Figure~\ref{fig:ppphenix} with a wide  $p_{T}$ range (5  $< p_{T} <$  65 GeV) are reconstructed via a gaussian filtering algorithm with a parameter of $\sigma=0.3$ at the PHENIX detector. This cross-section is compared to jets reconstructed with a cone radius of 0.4  from STAR at the same center of mass energy shown as magenta circles  \cite{starpp}.  Both jet cross-sections agree well with each other within their statistical and systematic  uncertainties. The STAR cone jets were compared to NLO pQCD cross-section using the CTEQ6M parton distributions \cite{starpp}. These calculations (also shown in Figure~\ref{fig:ppphenix} as the solid line) show a satisfactory agreement for cross-sections over 7 orders of magnitude \cite{starpp,phenixnlo}.  New measurements from STAR experiment for the sequential recombination algorithm jets using FastJet suite of algorithms are presented  in Figure~\label{fig:pp}  \cite{ploskon, fastjet}.  Resolution parameter of 0.4 is used for both $\rm k_{T}$ and anti-$\rm k_{T}$ algorithms and these cross-sections also agree well within the uncertainties with the previously published cone jets shown as stars in Figure~\ref{fig:pp} reconstructed with the STAR experiment \cite{starpp}.  

Jet measurements require various levels of corrections depending on the detector subsystems and jet algorithms response before they can be compared to theoretical calculations. Many of the preliminary estimation of these corrections are applied to the jets reconstructed in p+p collisions shown in Figures~\ref{fig:ppphenix} and \ref{fig:pp}. In the following sub-sections we review the main corrections for the jet measurements. We will concentrate this discussion with the STAR detector system only. 

\subsection{Particle and Jet Level Corrections}

Jets are reconstructed using cone and sequential recombination algorithms that cluster reconstructed time projection chamber (TPC) tracks and barrel electromagnetic calorimeter (BEMC) energy deposits.  Corrections for double-counting of energy due to hadronic energy deposition in the BEMC  of minimum ionizing particles and electrons is required at the particle level before running the jet finders to estimate both the jet energy and the underlying diffuse background of events correctly \cite{staret}.  Limited efficiency of tracking and unobserved neutral energy  such as $K^{0}_L$ mesons and neutrons will result into an energy shift and is required to be corrected at the jet level with Monte-Carlo studies and with data driven ways for example by utilizing the measured protons and charged kaons.  Similarly BEMC calibration scale is required to be assessed and corrected as it also results into an energy shift in the jet spectrum. It is possible to deconvolve the resulting gaussian smearing  purely due to detector effects such as TPC momentum and BEMC energy resolution from the jet distributions \cite{bhap}.   Fake jets that are purely due to random clustering of event fluctuations have to be identified and removed from jet spectrum statistically.  Detailed studies to estimate the fake jets with randomization of particles in an event and di-jets leads to similar results with the major contribution of fake jets for jets with a $p_{T}$ below 15 GeV  in the heavy-ion collisions \cite{brunawwnd,ploskon}.

\subsection{Underlying Heavy Ion Event Background}

The complex heavy ion background makes the full jet reconstruction a challenging task. Assumptions required by the full jet reconstruction in these complex environments require further investigation of the algorithmic responses of full jet reconstruction by utilizing various reconstruction algorithms.  A fundamental requirement that the signal and the background are two separable components can be violated by the presence of the jets and their effect on the background estimation. For example the initial state radiation, even though expected to be small compared to jet energy, might be different in A+A relative to p+p.  

Initial state processes resulting in the enhancement of the multiplicity of the underlying events appears to be distributed uniformly in the simpler p+p case \cite{helen} and therefore it is fully accounted for the estimation of the background under jets \cite{catchment, caccarinew}. However in the A+A collisions, ``the p+p correspondent" underlying event may be modified, possibly generating non-uniform structures. These non-uniformities in the background might be even larger  due to the final state processes.  Energy loss of the jet in matter might modify the event shape, resulting in non-uniform structures such as the ridge. 

Azimuthal and longitudinal anisotropy of heavy ion events will also result into non-uniform backgrounds.   Some of these sources of correlated backgrounds can be brought under quantitative control by using different collision systems. On the other hand, other observed effects might help us to understand details of the jet interactions with the heavy ion environment and may give further insight into the structures that are observed in di-hadron correlations and their origins.

\subsection{Biases or Physics Observables}

The ultimate goal of full jet reconstruction is to investigate the jet quenching in heavy ion collisions at the partonic level, without any ambiguities being introduced by hadronization and geometric biases of the inclusive spectrum and di-hadron measurements.  However,  it is possible that  new biases can be introduced when reconstructing jets. For example, all jet algorithms have various parameters for searching and defining jets, and the effects of varying these parameters need to be explored  in detail for a full understanding of jet reconstruction and for a systematic study of jet broadening effects.

A bias will be introduced while trying to reduce the effect of the background fluctuations in heavy ion collisions with the minimum threshold cuts on the track momenta and BEMC tower energies ($p_{T}^{cut}$).  Comparison of the jet spectra with a variation of $p^{min}_{T}$ threshold cuts in Au+Au and the $\rm N_{Binary}$ scaled p+p collisions show a good agreement between Au+Au and $\rm N_{Binary}$ scaled p+p  jet measurements for the lowest value of the $p^{min}_{T}$ cut but a much poorer one with the larger
$p_{T}$ threshold cut \cite{salurwwnd2, me}. This suggests that the threshold cuts introduce biases which are not fully corrected by the procedures that use fragmentation models that are developed for $\rm e^{+} + e^{-}$ and p+p collisions.

The resolution parameter or cone size, which restricts the area of the jet and thereby the amount of energy flow, can be a harder parameter to calculate hence interpret in heavy ion collisions than in p+p collisions.  If the jets are broader  in the heavy ion environment, the same resolution parameter might not be sufficient to recover the same fraction of jet energy in comparison to p+p jets.  This bias needs to be investigated by varying the resolution parameter and by looking into the jet profile of these jet definitions.   Variation of these biases introduced by both resolution parameter and  $p^{min}_{T}$ cut might serve useful to study the evaluation of the quenching on jets to determine the distribution of the lost energy \cite{vitev2}.

Another bias is introduced with the event selection. For example to enhance the recorded rate of high $p_{T}$ particles and jets, events above some threshold in the BEMC are collected. (This threshold is 5.4 GeV for the STAR experiment during 2006 and 2007 runs.)   This is very similar to the case of jets that are reconstructed with seeded infrared unsafe algorithms.  The online BEMC tower triggers introduce a strong bias of reconstructed jets that are fragmenting hard in comparison to the jets that are reconstructed without a seed.  New observables such as intra-jet energy distributions, away side of the tower triggered di-jets and hadron-jet correlation studies can be used to further elucidate the effect of these biases.

\section{Jet  Results in Heavy Ion Collisions}

\begin{figure}[h]
\centering
\begin{minipage}{16.5pc}
\includegraphics[width=16.5pc]{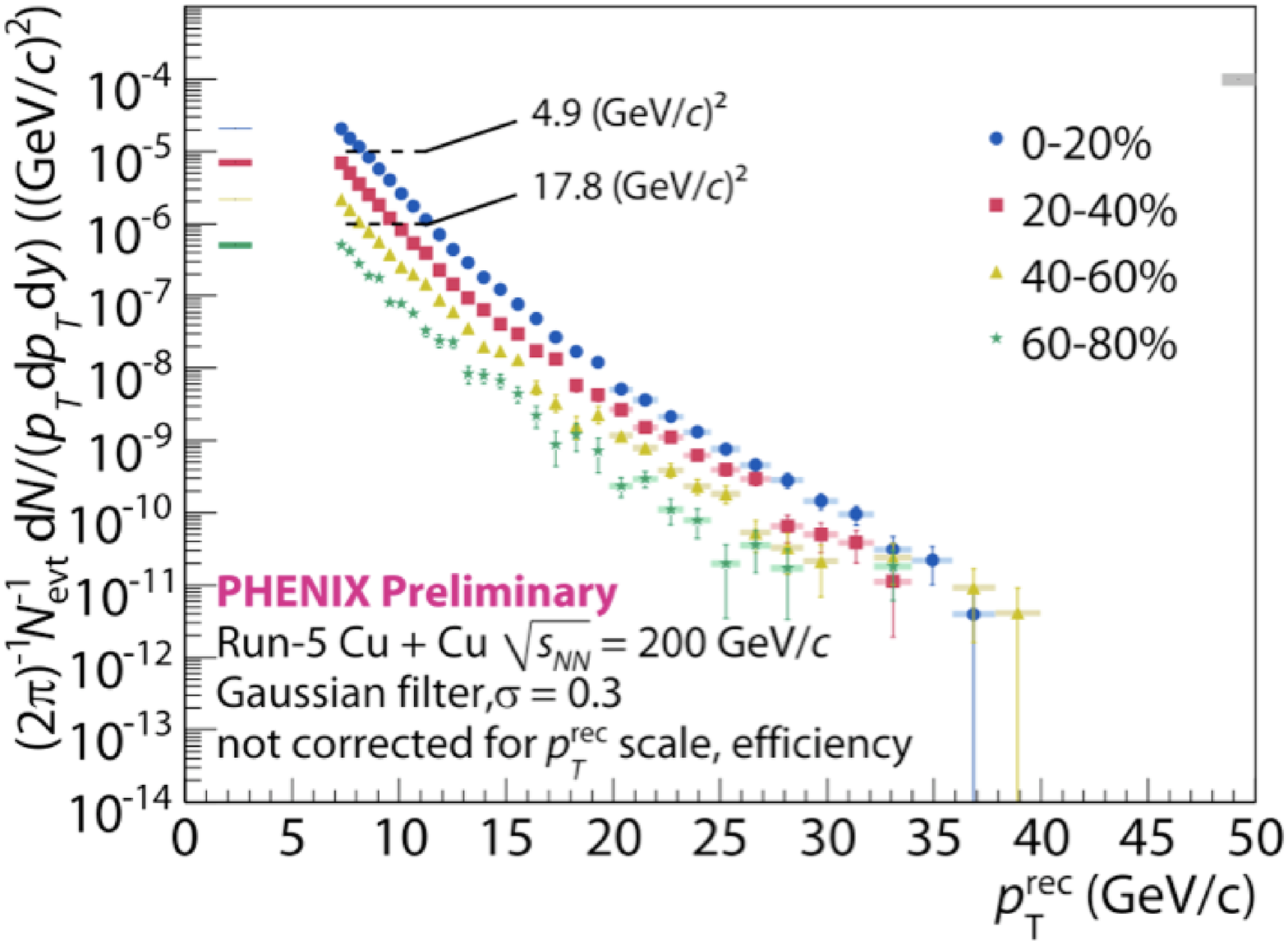}
\caption{\label{fig:cucu}  Raw jet yield vs jet $p_{T}$ obtained by the gaussian filtering algorithm after fake jet correction for $0-20\%$,  $20-40\%$,  $40-60\%$ and $60-80\%$ centralities of  Cu+Cu collisions \cite{yuiQM}.}
\end{minipage}\hspace{2pc}%
\begin{minipage}{15.5pc}
\includegraphics[width=15.5pc]{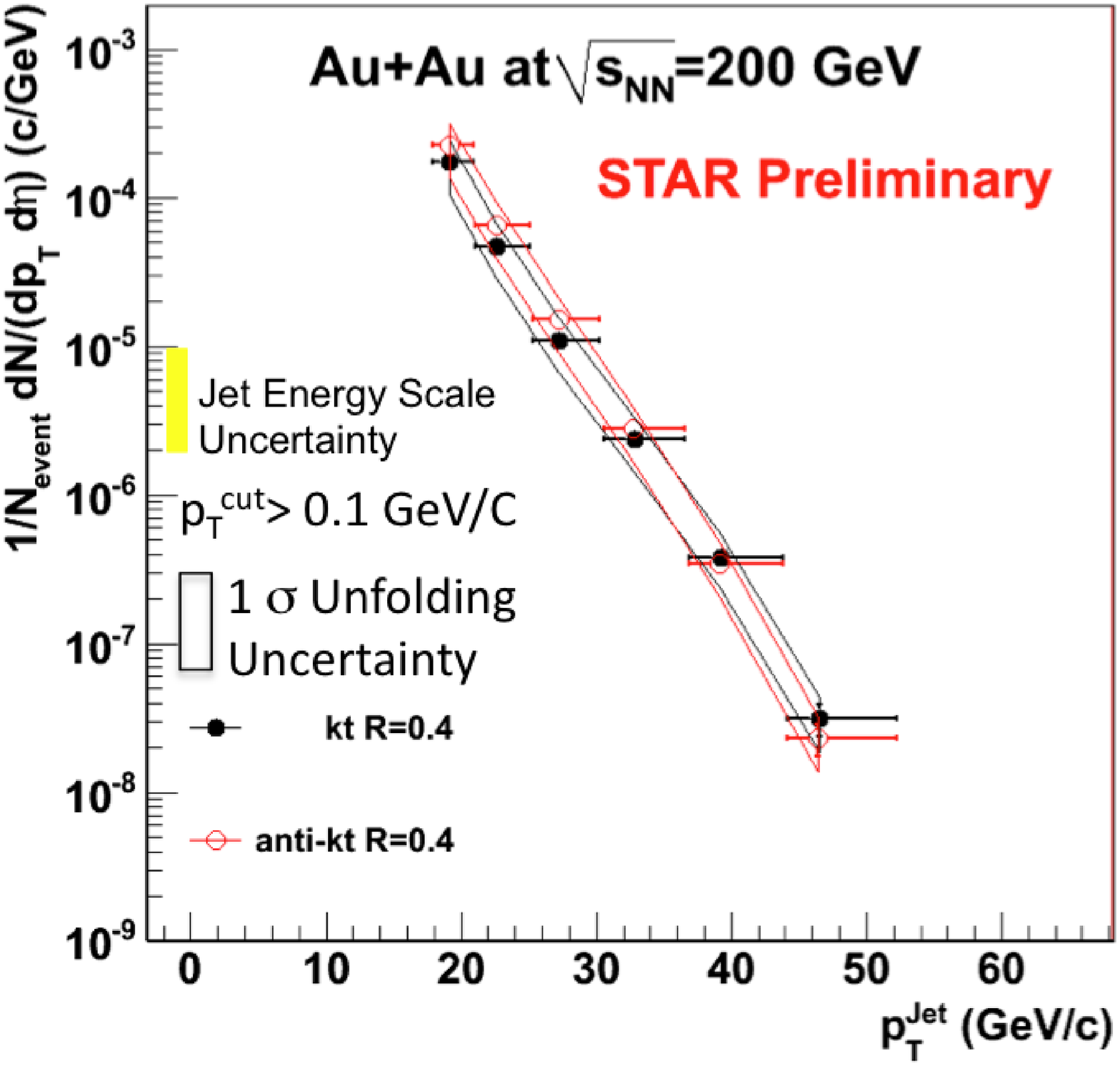}
\caption{\label{fig:kt} Jet yield per event vs transverse jet energy for the central Au+Au collisions obtained by the sequential recombination ($\rm k_{T}$ and anti-$\rm k_{T}$) algorithms from STAR .}
\end{minipage} 
\end{figure}

Uncorrected jet spectra reconstructed via a gaussian filtering algorithm in four centrality classes of Cu+Cu collisions at $\sqrt{s_{NN}}=200$ GeV collected by the PHENIX experiment is shown in Figure~\ref{fig:cucu} \cite{yuiQM}.  Unfolded and corrected jet results are presented later in \cite{yuidnp}. 
The Figure~\ref{fig:kt} shows the comparison of the inclusive jet spectra reconstructed by $\rm k_{T}$ and anti-$\rm k_{T}$ sequential recombination algorithms for the most central Au+Au collisions collected by the STAR experiment \cite{ploskon}.  Systematic uncertainties due to the unfolding procedure  are shown as the envelopes in red and black and the jet energy resolution as the yellow bar.  Jet spectra are consistent between the two  sequential reconstruction algorithms extending to 50 GeV kinematic reach.

 For the full jet reconstruction in heavy ion collisions, $\rm N_{Binary}$ scaling as calculated by a Glauber model \cite{glauber} is expected if the reconstruction is unbiased, i.e. if the full jet energy is recovered independent of the fragmentation details, even in the presence of strong jet quenching. This scaling is analogous to the cross section scaling of high $p_{T}$ direct photon production in heavy ion collisions observed by the PHENIX experiment \cite{phenix}.  
 
\begin{figure}[h]
\centering
\begin{minipage}{17pc}

\includegraphics[width=17.pc]{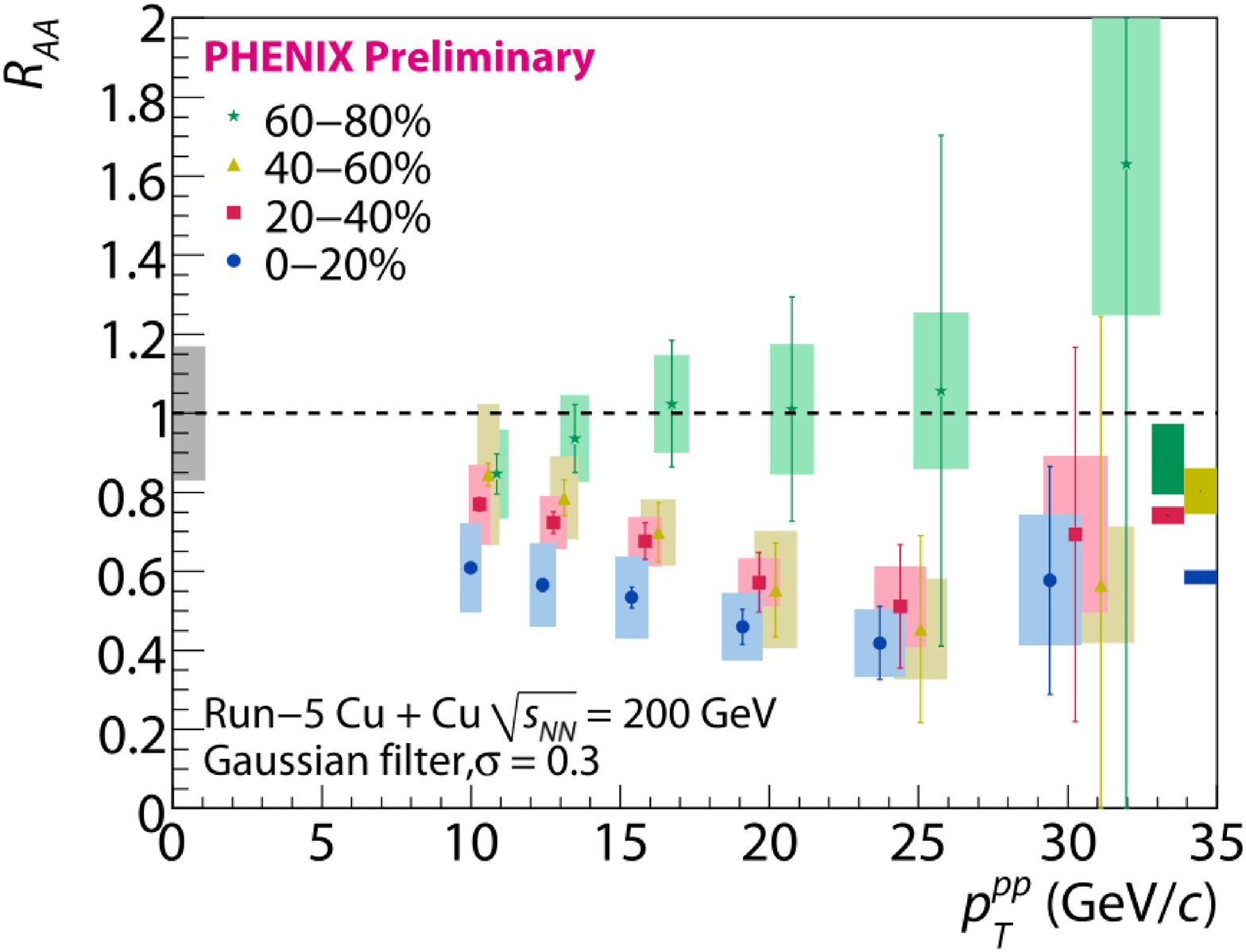}
\caption{\label{fig:raaphenix} Centrality dependence of the nuclear modification factors of reconstructed  jets (Cu+Cu collisions divided by $N_{Binary}$ scaled p+p collisions at $\sqrt{s_{NN}}=200$ GeV) measured with the PHENIX detector \cite{yuidnp}.  } 
\end{minipage}\hspace{2pc}%
\begin{minipage}{17.5pc}
\includegraphics[width=17.5pc]{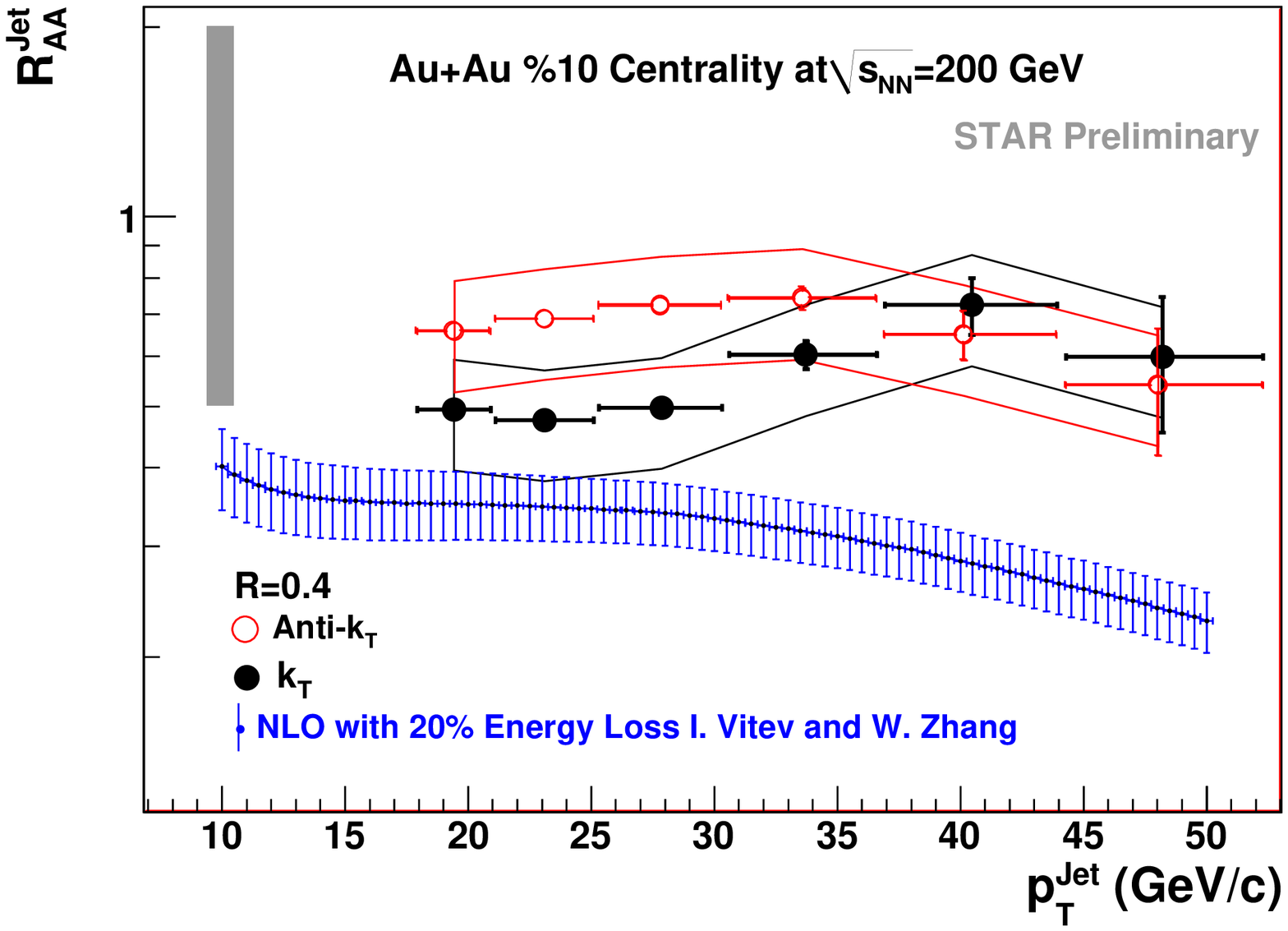}
\caption{\label{fig:raastar} Momentum dependence of the nuclear modification factor of jet spectra reconstructed with $ \rm k_{T}$ and anti-$\rm k_{T}$ algorithms (0-10\% most central Au+Au divided by $\rm N_{Binary}$ scaled p+p collisions) at the STAR detector and compared with a partonic level NLO calculation \cite{ploskon,vitevppr}. }

\end{minipage} 
\end{figure}

 The nuclear modification factor ($\rm R_{AA}$) for the jets defined with a gaussian filtering algorithm in the smaller system sizes such as Cu+Cu collisions are presented in Figure~\ref{fig:raaphenix}. While the experimental uncertainties are large, a trend towards a larger suppression reaching that of single particle suppression for higher centralities is observed.  This much suppression implies that the gaussian filtering algorithm with $\sigma=0.3$ seems to be geometrically  biased like the di-hadron correlations and is only recovering a small fraction of the jet energy rather than the full jet energy.

 The nuclear modification factor for the reconstructed jet spectra with the STAR detector for the most central Au+Au collisions at $\sqrt{s_{NN}}=200$ GeV can be found in Figure~\ref{fig:raastar}.  The envelopes shown around the data points for the resolution parameter of 0.4 from $\rm k_{T}$  and anti-$\rm k_{T}$ jets represent the one sigma uncertainty of the deconvolution of the heavy ion background fluctuations. The total systematic uncertainty on the jet energy scale is around 50\%, shown as the gray band.  Since the $\rm R_{AA}$ measurements are from two separate years main uncertainty  due to the BEMC calibration do not cancel as naively expected from the ratio.   How to improve the jet energy resolution at STAR is currently being explored.  The momentum dependence of the nuclear modification factor is  observed to be larger than the amount of suppression of $\pi$ meson $\rm R_{AA}$.  A hint of a suppression of jet $\rm R_{AA}$ below 1 for jets with $p_{T}$ above 30 GeV can also be observed with the large uncertainties in mind. It is expected that for a smaller resolution parameter in the sequential recombination algorithms, this suppression should reach to single particle suppression at large momentum.

 The systematic  variation of the size of the resolution parameter and the suppression amounts in nuclear modification factors can be used to evaluate the quenching effects and the expected amount of jet broadening. The jet $\rm R_{AA}$ of STAR is compared with next-to-leading order  calculation by I. Vitev and B.W. Zhang as presented by the blue histogram in Figure~\label{fig:raa} \cite{vitevppr}. Error bars around the theoretical data points are due to variation of the energy loss by 20\%. Details of these partonic level calculations and system size and resolution parameter dependences can be found in \cite{vitevppr}.    While the hadronization effect is expected to be small in a nuclear modification factor, STAR data points are systematically higher than that of the calculation. Improvements in the jet energy scale from the experimental side and  the addition of hadronization effects from the theoretical side  is required to investigate further if there is a real discrepancy in data and model calculation due to other unexplored quenching effects.

 \section{Conclusions}
 
Large momenta reaching 50 GeV can be studied in heavy ion collisions for the first time with the full jet reconstruction. New physics effects should be considered when interpreting the results at large momentum.  For example the momentum dependence of the relative contributions of quark and gluon sub-processes to inclusive jet production might be altered. The quark and gluon contributions already vary with respect to the jet momentum in elementary p+p collisions \cite{vogelsang}.  The relative contributions might be even more different in a heavy ion environment when a quark gluon plasma is produced, affecting the expected shape of the jet spectra and therefore of the nuclear modification factors. Another possibility is that at large momentum fraction $x$, initial state effects (such as the EMC effect which is the deviation between structure functions of heavy ions to light ions) are observed to be as large as 15\% \cite{emc}.  Some other contributions like the EMC effect might be playing a major role in the relative suppression or enhancement of nuclear modification factors at large momentum.

The new Monte-Carlo based simulations of  jet quenching in medium such as Jewel \cite{jewel}, Q-Pythia \cite{qpythia} and YaJEM \cite{yajem} and complementary analytic  calculations \cite{vitev2,vitevppr, vitev, borghini}  recently became  available to pursue a quantitative analysis of jet quenching as observed in heavy ion collisions.  However there are many uncertainties (e.g., how hadronization is treated) in the predictions  of these models and calculations.  To confront the calculations with data, new robust QCD jet observables that are unaffected by the $p_{T}$ cuts and hadronization need to be explored experimentally. For example the  subjet observable is infrared safe and insensitive to hadronization and will be used to study the jet quenching \cite{jewel}.

The studies shown here indicate that reconstruction of jets with a uniquely large kinematic limit is possible in heavy ion events. Jet reconstruction in heavy ion collisions is not yet free of biases due to selection of particles such as $p_{T}$ cuts to reduce the fluctuations of heavy ion background, requirement of algorithmic parameters such as cone size, gaussian width or resolution parameter, and the collection of events with thresholds to enhance the jet rates. Multiple channels like di-jets, h-jets, gamma-jets to measure qualitatively new observables such as energy flow, jet substructure and fragmentation functions will help to to assess fully the systematic uncertainties of jet measurements and the scale of jet broadening.

\section*{Acknowledgments}
The author wishes to thank Ivan Vitev for the valuable discussions and providing the theoretical calculations  and the organizers for the invitation and for the fruitful meeting.

\end{document}